\documentclass[9pt,journal,compsoc]{IEEEtran}
\ifCLASSOPTIONcompsoc
  \usepackage[nocompress]{cite}
\else
  \usepackage{cite}
\fi
\usepackage{graphicx}
  \usepackage{color}
  \usepackage{caption}
\usepackage{subcaption}
\usepackage[table,xcdraw]{xcolor}

\ifCLASSINFOpdf
\else
\fi

\begin{document}

%
%
%
%
%
%
%
%
%
%
%
%
%
%

%

\title{Power and Accuracy of Multi-Layer Perceptrons (MLPs)\\under Reduced-voltage FPGA BRAMs Operation}


%
%
%
%

\author{Behzad~Salami, 
        Osman~S.~Unsal, 
        and~Adrian~Cristal~Kestelman\\
\textit{BSC}}
\IEEEtitleabstractindextext{%
\begin{abstract}
In this paper, we exploit the aggressive supply voltage underscaling technique in Block RAMs (BRAMs) of Field Programmable Gate Arrays (FPGAs) to improve the energy efficiency of Multi-Layer Perceptrons (MLPs). Additionally, we evaluate and improve the resilience of this accelerator. Through experiments on several representative FPGA fabrics, we observe that until a minimum safe voltage level, \textit{i.e.,} $V_{min}$ the MLP accuracy is not affected. This safe region involves a large voltage guardband. Also, it involves a narrower voltage region where faults start to appear in memories due to the increased circuit delay, but these faults are masked by MLP, and thus, its accuracy is not affected. However, further undervolting causes significant accuracy loss as a result of the fast-increasing high fault rates. Based on the characterization of these undervolting faults, we propose fault mitigation techniques that can effectively improve the resilience behavior of such accelerator. Our evaluation is based on four FPGA platforms. On average, we achieve $>$90\% energy saving with a negligible accuracy loss of up to 0.1\%.     
\end{abstract}

}

\maketitle

\IEEEdisplaynontitleabstractindextext

%
\IEEEpeerreviewmaketitle

\IEEEraisesectionheading{\section{Introduction}\label{sec:introduction}}

%
%
%
%
\IEEEPARstart{F}{PGAs} are continually obtaining more attention to accelerate state-of-the-art applications \cite{salami4, salami5, salami6, salami8, salami15} like Neural Networks (NNs) \cite{Guo}, thanks to their massively parallel architecture, data-flow execution model, and reconfigurability feature as well as the recent advances on High-Level Synthesis (HLS) tools. However, the energy efficiency of such accelerators is still a key concern, reported to be at least one order of magnitude less than customized Application-Specific Integrated Circuit (ASIC)-based models \cite{Eriko}. To bridge this gap, several design-, compile-, and application-level techniques can be applied to FPGAs. As an orthogonal hardware-level approach, in this paper, we propose to utilize aggressive supply voltage underscaling. This technique can significantly improve the energy efficiency of the underlying hardware; however, as a downside, it may cause timing faults. In the NN applications including in Multi-Layer Perceptron (MLPs), these timing faults can, in turn, degrade the accuracy. In this paper, we experimentally evaluate the energy-accuracy trade-off of a typical FPGA-based MLP under extremely low-voltage operations of on-chip memories, \textit{i.e.} Block RAMs (BRAMs). We implement and demonstrate our undervolting technique on four representative FPGAs from Xilinx, a main vendor, to consider the FPGA-to-FPGA variation in the results.

\begin{figure}[!t]
\centering
\includegraphics[width=\linewidth]{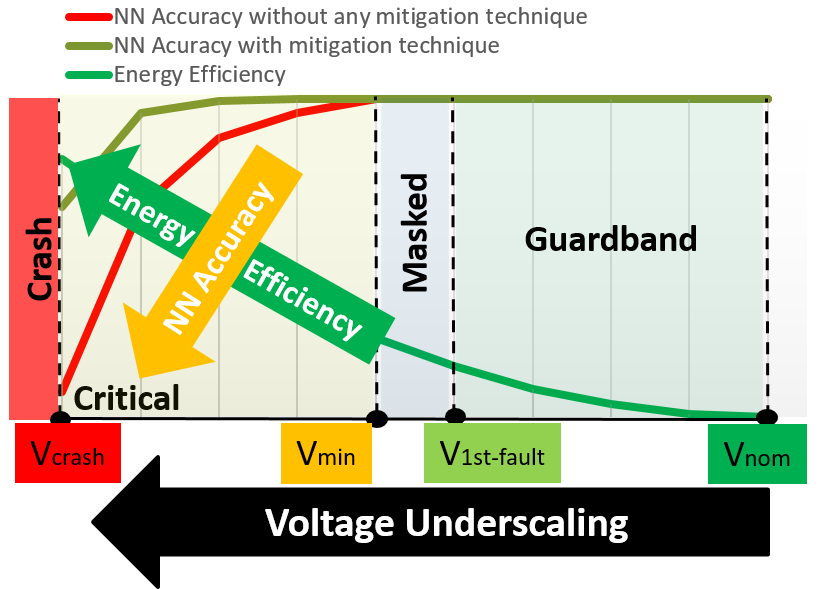}
\caption{The overall energy/accuracy trade-off. \\$V_{nom}$: The default voltage level. \\$V_{1st-fault}$: The voltage level that the first fault appears. \\$V_{min}$: Below this voltage level there is MLP accuracy loss. \\$V_{crash}$: Below this voltage level FPGA crashes.}
\label{fig:overall-voltage} 
\end{figure}
The overall voltage behavior observed is illustrated in Fig. \ref{fig:overall-voltage} and the description of subsequent voltage regions, \textit{i.e.,} \textbf{Guardband}, \textbf{Masked}, \textbf{Critical}, and \textbf{Crash} are summarized in Table. \ref{table:regions}. As seen, by voltage underscaling below the default level \textit{i.e.,} $V_{nom}$, there is a \textbf{Guardband} region. In this region, there is energy efficiency improvement without compromising the MLP accuracy or performance since no fault appears. By further undervolting below the guardband and due to the circuit delay path increase, faults start to appear at $V_{1st-fault}$. However, until a minimum safe voltage level, \textit{i.e.,} $V_{min}$, relatively lower fault rate is automatically covered by MLP and thus, there is no accuracy loss, \textit{i.e.,} \textbf{Masked} region. By further voltage underscaling below $V_{min}$, the MLP accuracy starts to being degraded, \textit{i.e.,} \textbf{Critical} region. For instance, we observe that decreasing the voltage by $50mV$ leads to an MLP accuracy loss of up to 3.46\%. To mitigate this accuracy loss, we propose effective fault mitigation techniques that can significantly prevent the accuracy loss up to 0.1\%. Also, thanks to our fault mitigation techniques, $V_{min}$ decreases, and thus, the MLP accuracy starts to be degraded in lower voltages of up to $30mV$. Finally, by further voltage underscaling, the FPGA system crashes with no response at $V_{crash}$, \textit{i.e.,} \textbf{Crash} region.

\begin{table}[]
\caption{Different voltage regions on $V_{CCBRAM}$.}
\label{table:regions}
\begin{tabular}{l|llll}
 & \begin{tabular}[c]{@{}l@{}}Voltage\\ Margin\end{tabular} & \begin{tabular}[c]{@{}l@{}}Faults\\ Appear?\end{tabular} & \begin{tabular}[c]{@{}l@{}}Accuracy\\ Loss?\end{tabular} & \begin{tabular}[c]{@{}l@{}}Action to\\ be taken\end{tabular} \\ \hline
\textbf{Guardband} & {[}$V_{nom}, V_{1st-fault}$) & No & No & ... \\
\textbf{Masked} & {[}$V_{1st-fault}, V_{min}$) & Yes & No & ... \\
\textbf{Critical} & {[}$V_{min}, V_{crash}$) & Yes & Yes & Mitigation \\
\textbf{Crash} & \textless{}$V_{crash}$ & ... & ... & ...
\end{tabular}
\end{table} 

\begin{figure*}
\centering
\begin{subfigure}{0.248\linewidth}
  \centering
  \includegraphics[width=\linewidth]{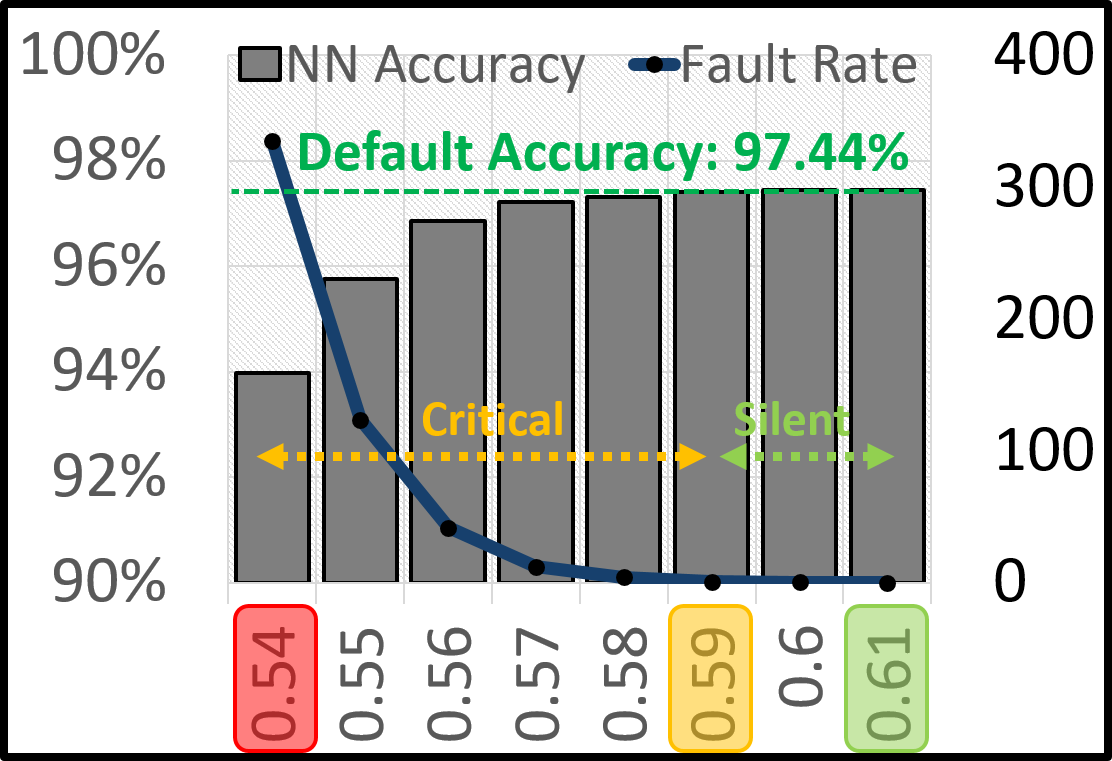}
  \caption{VC707.}
  \label{fig:loss-VC707}
\end{subfigure}%
\begin{subfigure}{0.248\linewidth}
  \centering
  \includegraphics[width=\linewidth]{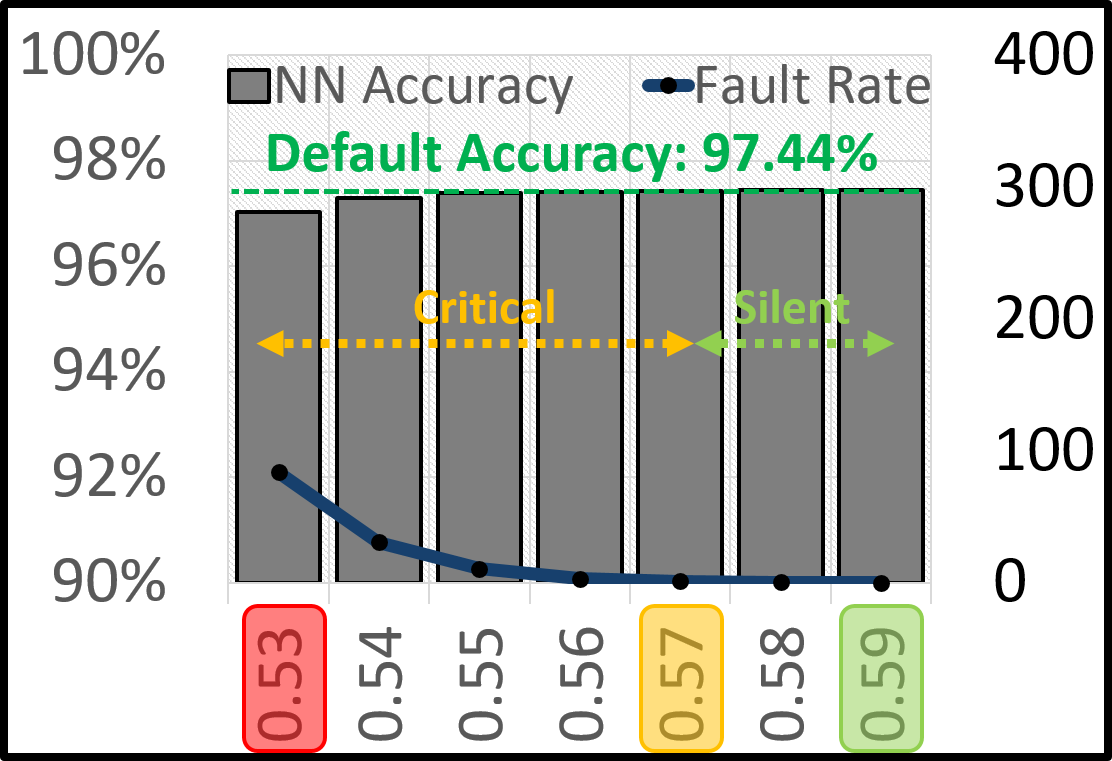}
  \caption{ZC702.}
  \label{fig:loss-ZC702}
\end{subfigure}
\begin{subfigure}{0.248\linewidth}
  \centering
  \includegraphics[width=\linewidth]{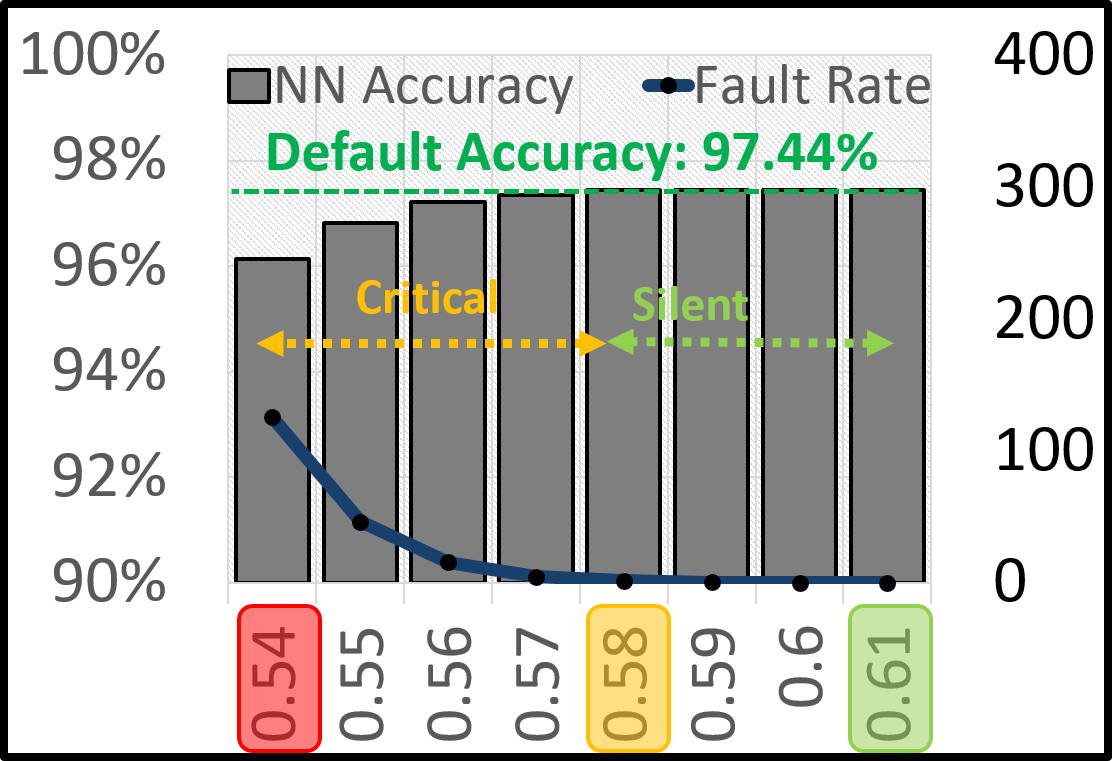}
  \caption{KC705-A.}
  \label{fig:loss-KC705A}
\end{subfigure}%
\begin{subfigure}{0.248\linewidth}
  \centering
  \includegraphics[width=\linewidth]{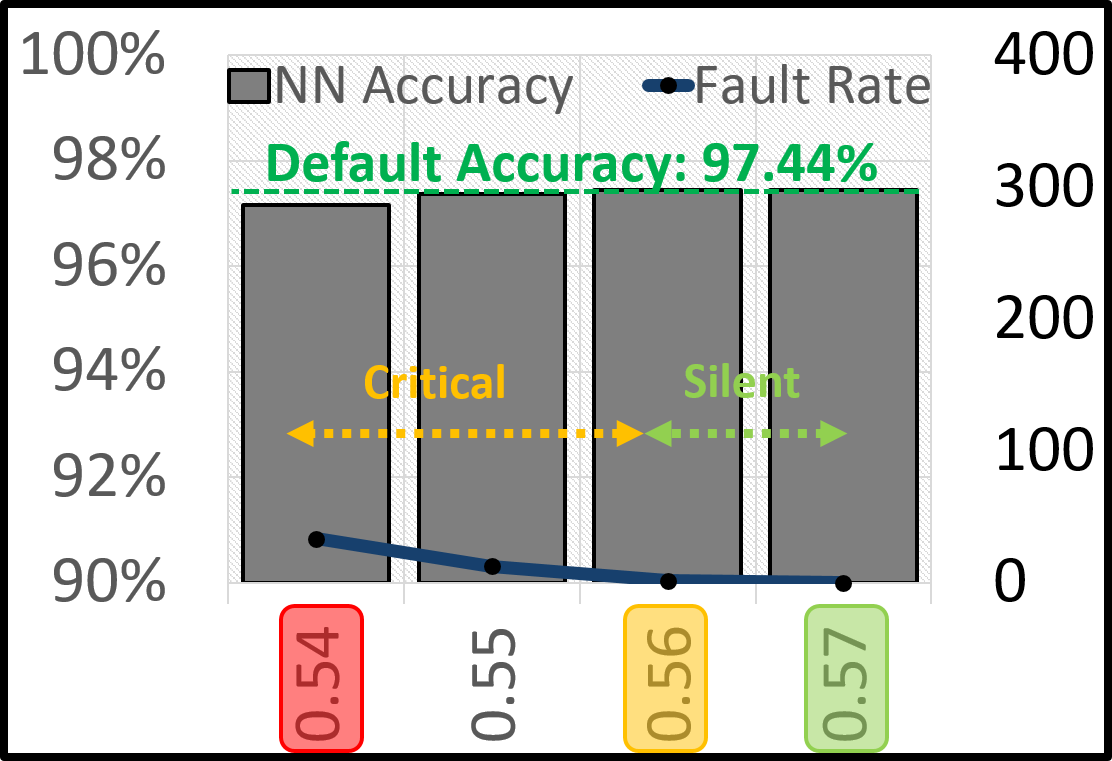}
  \caption{KC705-B.}
  \label{fig:loss-KC705B}
\end{subfigure}
\caption{Resilience behavior of the reduced-voltage MLP on four studied FPGAs (x-axis: $V_{CCBRAM}$ (V), y-axisL: MLP inference error rate (percentage), y-axisR: BRAMs fault rate (per 1Mb), shown for \textbf{Masked} [$V_{1st-fault}, V_{min}$) and \textbf{Critical} [$V_{min}, V_{crash}$) regions. \\ + \textcolor{green}{$V_{1st-fault}$}, \textcolor{orange}{$V_{min}$}, and \textcolor{red}{$V_{crash}$} are highlighted. \\ + Among different platforms, slight variation of the voltage regions and the subsequent significant impact on the fault rate and MLP accuracy in the \textbf{Critical} region can be seen.}
\label{fig:loss}
\end{figure*}

By experimenting on several representative FPGAs, we evaluate the FPGA-to-FPGA variation and observe that the different voltages, \textit{i.e.,} $V_{1st-fault}$, $V_{min}$, and $V_{crash}$ have slight variability; however, the fault rate and in turn, the MLP accuracy loss in the \textbf{Critical} region is significant, which can be the consequence of the process variation, architectural differences, or aging. This paper experimentally explores and evaluates different voltage regions of the FPGA-based MLP under aggressive low-voltage operations for on-chip memories and optimizes the energy-resilience trade-off by encapsulating the following contributions:
\begin{itemize}
    \item Improving the \textbf{ENERGY} efficiency of FPGA-based MLPs through aggressive voltage underscaling on BRAMs from the default level until the lowest possible level in which the system crashes. The energy saving gain is $>$90\% for on-chip memories.
    \item Improving the \textbf{RESILIENCE} of the FPGA-based MLPs below the safe voltage region. We propose efficient fault mitigation techniques to decrease the $V_{min}$ and to limit the accuracy loss to a maximum of 0.1\%. 
\end{itemize}

The rest of this paper is organized as follows. In Section \ref{sec:methodology}, we introduce the experimental methodology. In Section \ref{sec:FPGAVolt}, we present and discuss the energy-resilience trade-off results and fault mitigation techniques. We review the previous work in Section \ref{sec:Related}, and finally, the paper is summarized and concluded in Section \ref{sec:conclusion}.

\section{Experimental Methodology}\label{sec:methodology}
In this section, we briefly explain the model of the application as well as the undervolting methodology.
\subsection{Multi-Layer Perceptron (MLP) Model}
We perform our experiments on a fully-connected neural network, \textit{i.e.,} MLP, in the inference phase. It is a state-of-the-art model for small-medium size datasets, forms the most-computational part of Convolutional NNs (CNNs), and relatively less development has been made for them in comparison to CNNs \cite{Guo}. Our model is composed of input, hidden, and output layers, where all neurons of adjacent layers are fully connected. The intensity of each connection is determined by weights, whose values are tuned off-line in the training phase. Our tested MLP has a 6-layers topology, composed of one input, four hidden, and one output layer(s) with sizes of 784, 1024, 512, 256, 128, and 10 neurons, respectively. Also, for representing data, we use the fixed-point low-precision model. We evaluate this model on the MNIST dataset as a state-of-the-art image recognition benchmark \cite{mnist}. MNIST is a set of image with black and white digitized handwritten digits, each image is 784*8-bit pixels, and the output infers the number from 0 to 9. The training phase of the MLP is performed off-line using 60000 training images of MNIST on the software. Our model can reach to 97.44\% accuracy on 10000 MNIST inference dataset. 
 
 In the architecture of accelerator, weights of the MLP are located inside BRAMs, and the input images are being streamed through the off-chip DDR-3 memory. The required calculation of the image classification, \textit{i.e.}, matrix multiplication, and activation function are performed in parallel by leveraging DSPs and LUTs of the FPGA in a stream-fashion model, as typical \cite{Guo}. Our design utilizes more than 90\% of BRAMs, and its maximum working frequency is 200Mhz.

\subsection{Undervolting Below the Nominal Level ($V_{nom}$)}
We perform our undervolting experiments on several Xilinx FPGA platforms with 28nm technology, \textit{i.e.,} VC707, ZC702, and two identical samples of KC705, representing performance-oriented, software/hardware co-design, and power-optimized designs, respectively. Also, among different FPGA components, we concentrate on BRAMs, since first, they play a key role in the structure of the accelerator to locate the MLP weights on-chip; second, unlike other FPGA components, they have an independent voltage rail in the studied FPGA platforms, \textit{i.e.,} $V_{CCBRAM}$. BRAMs are small memory blocks that are distributed over the chip, and each basic BRAM block is a matrix of bitcells composed of rows and columns. In studied platforms, the size of each basic setup BRAM is 18 Kbits with 1024 rows and 18 columns. The default/nominal voltage of BRAMs, \textit{i.e.,} $V_{nom}$ is 1V for all of the studied platforms, set by the vendor. For the voltage scaling, we use Power Management Bus (PMBus) standard to access the on-board voltage regulator. We underscale the supply voltage by the scale of $10mV$. Finally, we report the total power consumption, including dynamic and static parts, measured using PMBus. 

\section{Experiemntal Results}\label{sec:FPGAVolt}
In this section, first, we discuss different voltage regions explored via undervolting, and second, we present our fault mitigation techniques in the Critical region where due to high fault rates the MLP accuracy loss is significant.

\subsection{Different voltage Regions}
As described earlier in Section \ref{sec:introduction}, we observe four voltage regions. Among studied platforms, there is a slight variability in the size of these regions, as detailed in Fig. \ref{fig:loss}. However, as seen, the fault rate in the \textbf{Critical} region and the subsequent impact on the MLP accuracy is significantly different among the studied platforms. This variability can be the result of the process variation, aging, or architectural differences among them. Below, we describe them in detail:
\begin{itemize}
    \item \textbf{Guardband Region:}  By voltage underscaling of $V_{CCBRAM}$ below $V_{nom}= 1V$, we observe a large voltage guardband in [$V_{nom}$ and $V_{1st-fault}$) for all platforms. The size of the Guardband region is measured to be $405mV$ on average. Guardbands are set by vendors to guarantee the worst-case circuit and environmental conditions. In this voltage region, there is no fault in BRAMs, and subsequently, there is no MLP accuracy loss.
    \item \textbf{Masked Region:} By further voltage underscaling below $V_{1st-fault}$ and until $V_{min}$, faults start to appear in BRAMs; however, the MLP accuracy is not affected, which can be due to the inherent robustness of the MLP for low fault rates. In other words, faults occur in this region but are masked by the MLP. The size of this area is measured to be $20mV$ on average. We observe that our design is inherently robust to 1.4 faults/Mbit that occurs at $V_{min}= 575mV$, on average across all studied platforms.   
    \item  \textbf{Critical Region:} By further voltage underscaling below $V_{min}$, the fault rate fastly increases and subsequently, the MLP starts to lose the accuracy. As shown in Fig. \ref{fig:loss}, there is a significant variation of the fault rate and thus, accuracy loss among platforms. For instance, as the best/worst platform, the voltage underscaling from $V_{min}= 0.59V/0.56V$ to $V_{crash}= 0.54V/0.54V$ in VC707/KC705-B causes 334.7/36.9 faults/1Mbit and 3.46\%/0.28\% MLP accuracy loss. To prevent this accuracy loss, our design is equipped with effective mitigation techniques that are discussed in Section \ref{subsec:mitigation}.
    \item \textbf{Crash Region:} Finally, the system crashes below the $V_{crash}$, and there is no response from FPGA platforms. $V_{crash}$ is the lowest voltage level that we could practically underscale. We measure it to be on average of $535mV$ with a slight variability among platforms.   
\end{itemize}

\begin{figure}[!t]
\centering
\includegraphics[width=\linewidth]{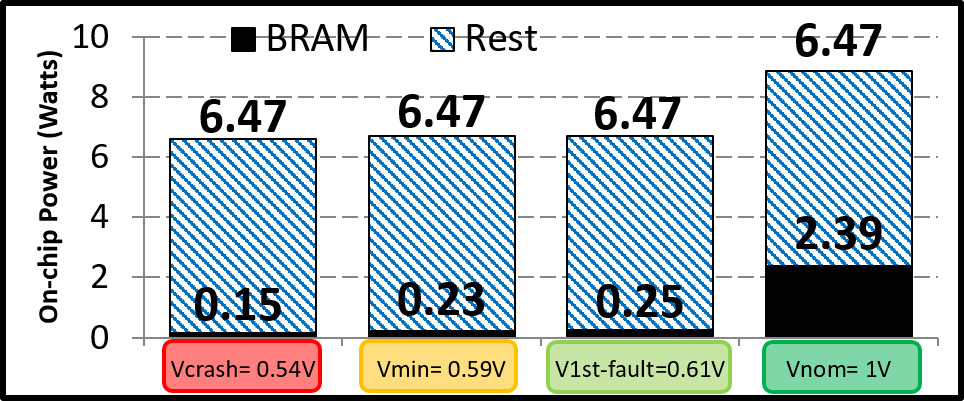}
\caption{Power saving at different voltage regions, shown for VC707 (similar for other platforms).}
\label{fig:breakdown} 
\end{figure}

By voltage underscaling, the power consumption and in turn, the energy dissipation also gradually decrease, as shown in detail in Fig. \ref{fig:breakdown} for VC707. We achieve an average of more than 90\% of BRAMs power dissipation savings at $V_{crash}$ in comparison to the same design at $V_{nom}$. 

\subsection{Fault Mitigation Techniques}
\label{subsec:mitigation}
As mentioned earlier, there is a significant accuracy loss when $V_{CCBRAM}$ is underscaled below the $V_{min}$. Relying on the behavior of the undervolting faults, we present techniques to mitigate the undervolting faults. These techniques, \textit{i)} prevents the MLP accuracy loss, and \textit{ii)} decreases the $V_{min}$ where MLP accuracy starts to be degraded. For instance, on VC707, our best technique can decrease the $V_{min}$, for $30mV$; also, it can limit the MLP accuracy loss to up to 0.1\% at $V_{crash}$. 

\begin{figure}[!t]
\centering
\includegraphics[width=\linewidth]{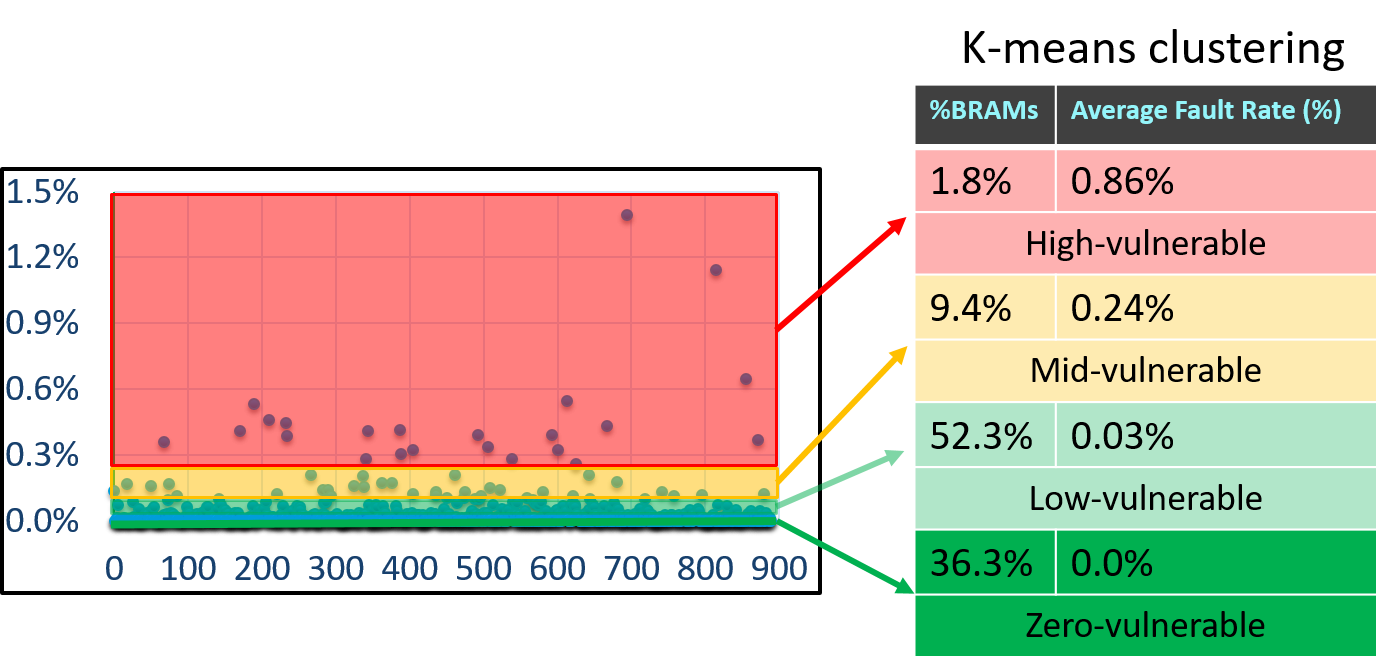}
\caption{Non-uniform fault distribution among BRAMs for VC707 with 2030 BRAMs, classified using the K-means clustering in terms of the fault rate at $V_{crash}$ (similar for other platforms).}
\label{fig:distribution} 
\end{figure}

\begin{figure}[!t]
\centering
\includegraphics[width=\linewidth]{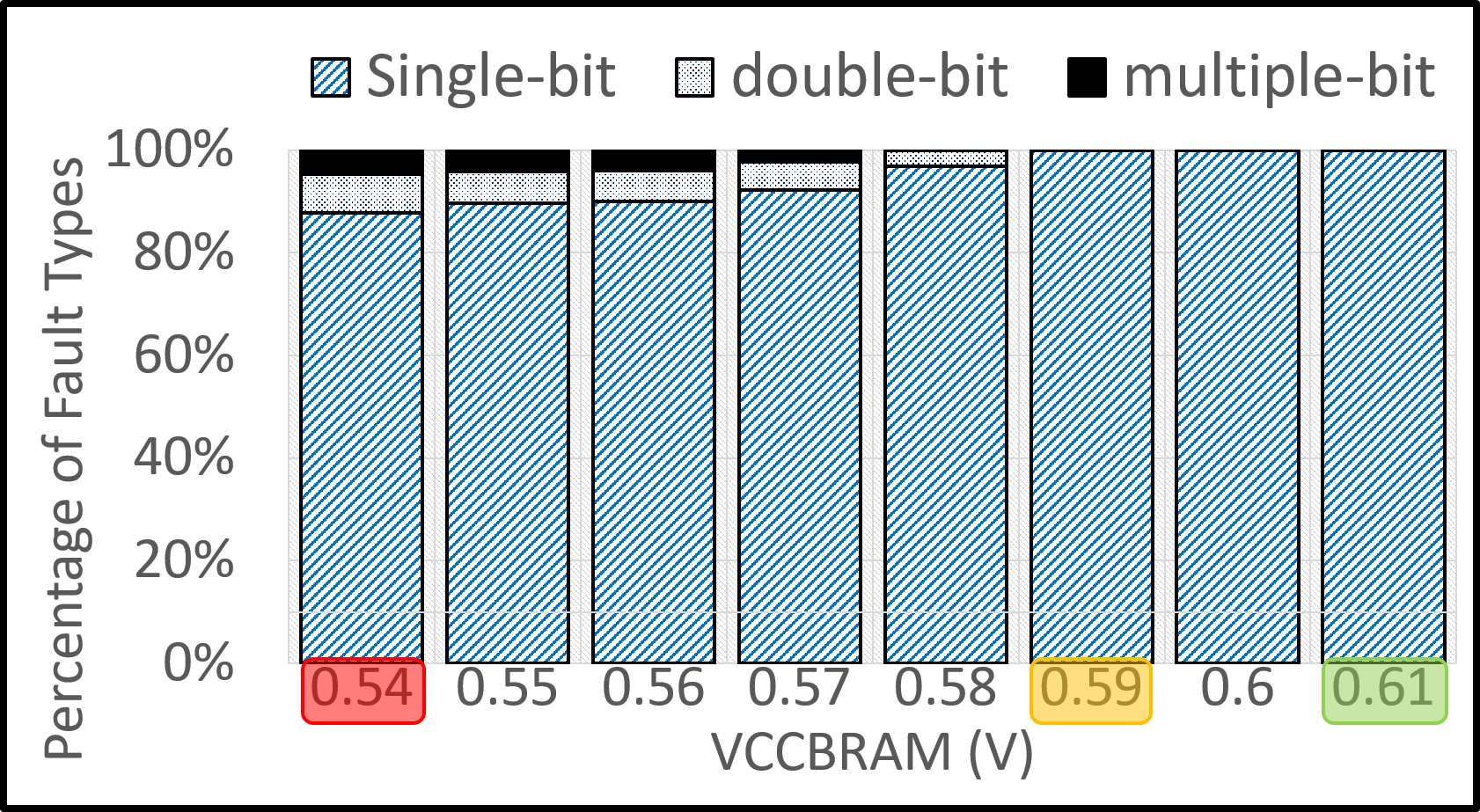}
\caption{Different types of undervolting faults, shown for VC707 (similar for other platforms).}
\label{fig:faultTypes} 
\end{figure}

\subsubsection{Intelligently-constraint Memory Mapping (IMM)}
By characterizing the undervolting faults, we observe that the faults are \textbf{fully non-uniformly} distributed among different BRAMs. For instance, as shown in Fig. \ref{fig:distribution} for VC707 at $V_{crash}$, only 1.8\% of BRAMs, tagged as High-vulnerable, experience a vast majority ($>$90\%) of faults. Keeping this point in the mind, Intelligently-constraint Memory Mapping (IMM) aims to eliminate High-vulnerable BRAMs. Toward this goal, IMM adds additional constraints for the Placement stage of the design compile using Physical Blocks (Pblocks) facility of Vivado, compile tool for Xilinx FPGAs. Note that due to a small percentage of High-vulnerable BRAMs, the timing slack overhead of the IMM is negligible. IMM shows significant efficiency to prevent the MLP accuracy loss; although, faults in non-High-vulnerable BRAMs still cause some accuracy loss of up to 0.85\% at $V_{crash}$, see Fig. \ref{fig:IMMECC}.    

\subsubsection{Error Correction Code (ECC)}
As another fault mitigation technique, we evaluate the built-in ECC of BRAMs. It is based on Hamming code with the type of Single-Error Correction and Double-Error Detection (SECDED), which can correct single-bit faults and detect (but not correct) double-bit faults. By an off-line fault characterization, we found that a vast majority ($\sim$ 90\% at $V_{crash}$ and even more in the higher voltage levels) of undervolting faults are \textbf{single-bit}, see Fig. \ref{fig:faultTypes}. The built-in SECDED-type ECC of BRAMs can efficiently mitigate most of these faults. Hence, we utilize ECC of BRAMs. As can be seen in Fig. \ref{fig:IMMECC}, the MLP accuracy loss is significantly prevented, and by voltage underscaling until $0.57V$ there is no effect on the MLP (without any mitigation, the $V_{min}$ is $0.59V$). However, due to those faults that ECC could not correct, \textit{i.e.}, double-bit, multiple-bit, and ECC-module corrupted faults, there is still some accuracy loss of up to 0.57\% at $V_{crash}$.  

\subsubsection{IMM+ECC}
\begin{figure}[!t]
\centering
\includegraphics[width=\linewidth]{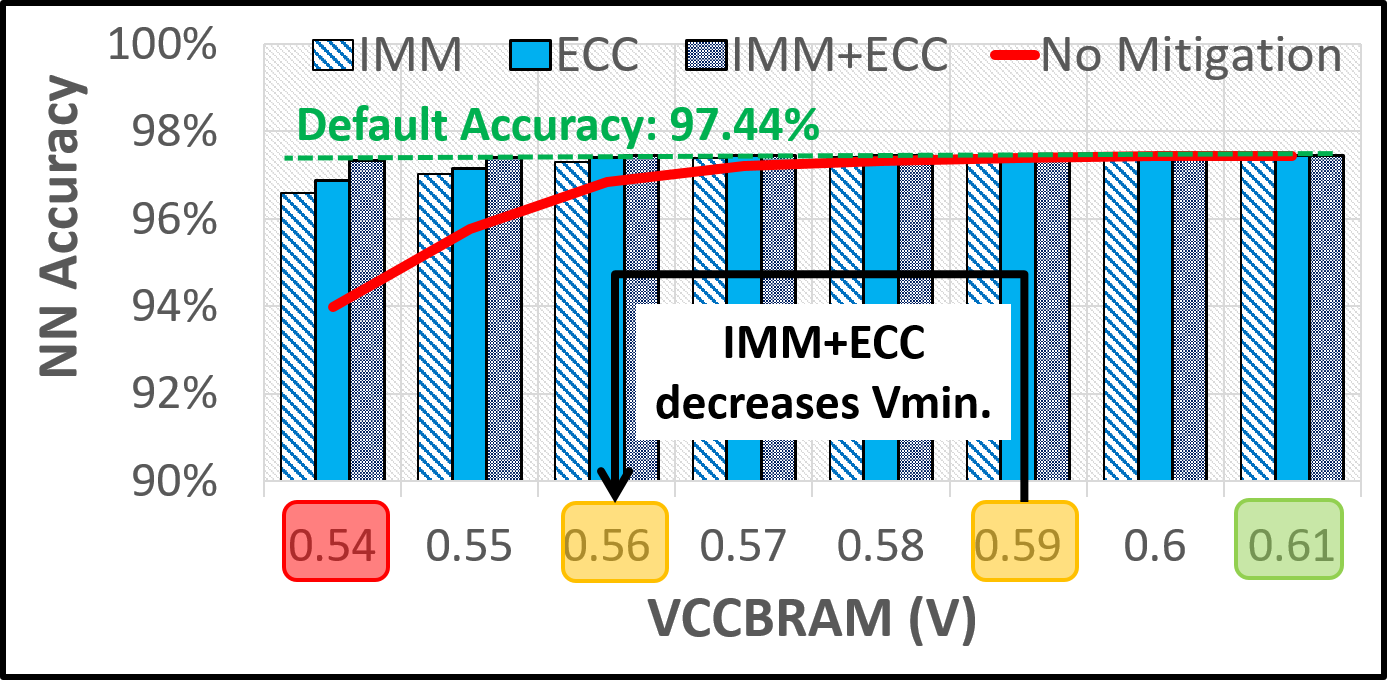}
\caption{Fault mitigation techniques, shown for VC707.}
\label{fig:IMMECC} 
\end{figure}

As mentioned earlier, IMM eliminates the High-vulnerable BRAMs; however, faults remained in other BRAMs can affect the MLP accuracy, as seen in Fig. \ref{fig:IMMECC}. On the other side, we observed that a vast majority of faults in other BRAMs are single-bit; thus, the built-in ECC can effectively cover them. In other words, we found ECC a useful complementary for IMM technique; \textit{i.e.,} ECC can cover those faults that are not covered by IMM. The combined IMM+ECC mitigation technique has a remarkable performance to cover the undervolting faults and in as shown in Fig. \ref{fig:IMMECC}, the $V_{min}$ is decreased for $30mV$ (from $0.59V$ to $0.56V$) and the MLP accuracy loss is limited to up to 0.1\% at $V_{crash}= 0.54V$ on VC707, \textit{i.e.}, the least-robust FPGA platform that we studied. 

\section{Related Work}\label{sec:Related}
\subsection{Undervolting}
Supply voltage underscaling has obtained significant attention and studied for modern CPUs \cite{Papadimitriou2}, GPUs \cite{An1}, and DRAMs \cite{dram}. Undervolting below the $V_{min}$, the most common approach was to prevent the fault using frequency underscaling \cite{GPUFreq}, which can limit the energy saving gain. Although, in part, other techniques like ECC has also been considered for processors \cite{itanium}. The related works on the FPGA voltage underscaling are usually accompanied by frequency underscaling \cite{FPGAFreq2}. More comprehensively, we evaluate the behavior of MLPs under reduced-voltage operations in FPGA BRAMs, explores and experimentally analyzes different voltage regions, and finally improves the previous and proposes novel mitigation techniques to cover undervolting faults. More details on our FPGA undervolting technique can be found in \cite{salami1, salami2, salami9, salami11, salami12, salami17, salami13, salami18}, being conducted under LEGaTO project \cite{salami7, salami10, salami16}.

\subsection{Resilience Studies on Neural Networks}
With technology scale developing, the resilience of NNs can be significantly affected due to the fabrication process uncertainties, soft-errors, harsh and noisy environments, aggressively low-voltage operations, among others. Hence, recently, the resilience of NNs has been studied in different abstraction levels. Most of these works are simulation-based efforts \cite{DNNResSW2, salami3, salami14}, in which, their verification on the real fabric can be a key concern. However, there are some efforts on real hardware too, mostly on ASICs \cite{DNNResHW2}. We complement the previous studies by experimenting MLPs on extremely low-voltage COTS real FPGA fabrics, \textit{i.e.} undervolting fault characterization and mitigation.

\section{Conclusion}\label{sec:conclusion}
In this paper, we evaluate a reduced-voltage on BRAMs and fault-resilient FPGA-based MLP. Our design delivers on average $>$90\% of the energy efficiency (with a negligible 0.1\% of the accuracy loss) in comparison to the baseline FPGA design at the default voltage level. Our prototype is on four real FPGA fabrics to evaluate the FPGA-to-FPGA variation experimentally. We characterize the effect of the reduced-voltage operations on the NN accuracy and accordingly, categorize different voltage regions. To alleviate the accuracy loss issues below the safe voltage region, our design is equipped with efficient techniques which rely on the behavior of undervolting faults. These techniques effectively prevent accuracy loss and decrease the $V_{min}$, \textit{i.e.,} minimum safe voltage level. 

\section*{Acknowledgments}
The research leading to these results has received funding from the European Union’s Horizon 2020 Programme under the LEGaTO Project (www.legato-project.eu), grant agreement No. 780681.

\end{document}